\documentclass[prl,twocolumn,superscriptaddress,preprintnumbers,aps,showpacs,amsmath,amssymb,reprint]{revtex4}

\usepackage{graphicx}
\usepackage{dcolumn}
\usepackage{bm}
\begin{document}

\title{General Criteria for Determining Rotation or Oscillation\\ in a Two-dimensional Axisymmetric System}

\author{Yuki Koyano}
\affiliation{Department of Physics, Graduate School of Science, Chiba University, Chiba 263-8522, Japan}

\author{Natsuhiko Yoshinaga}

\affiliation{WPI-Advanced Institute for Materials Research, Tohoku University, Sendai 980-8577, Japan}

\author{Hiroyuki Kitahata\footnote{Corresponding author. E-mail: kitahata@chiba-u.jp.}}

\affiliation{Department of Physics, Graduate School of Science, Chiba University, Chiba 263-8522, Japan}

\date{\today}% It is always \today, today,
             %  but any date may be explicitly specified

\begin{abstract}
A self-propelled particle in a two-dimensional axisymmetric system, such as a particle in a central force field or confined in a circular region, may show rotational or oscillatory motion.
These motions do not require asymmetry of the particle or the boundary, but arise through spontaneous symmetry breaking. 
We propose a generic model for a self-propelled particle in
 a two-dimensional axisymmetric system.
A weakly nonlinear analysis establishes criteria for determining rotational or
 oscillatory motion.
\end{abstract}

%\pacs{05.45.-a, 02.30.Oz, 45.50.-j, 05.45.Xt}

%\keywords{Suggested keywords}%Use showkeys class option if keyword
                              %display desired
\maketitle

\section{INTRODUCTION}

Motility is realized in nonequilibrium systems through energy gain and dissipation from/to the surroundings~\cite{Ramaswamy}. 
This is ubiquitous in biological systems, on scales from single cells to multicellular organisms. 
Such motion comes in many varieties; it may be ballistic, bidirectional, rotational, and so on~\cite{Badoual:2002, Berg, forest}.
Physico-chemical systems mimicking the motion of living things are thus useful for investigating the mechanical and statistical aspects of the motility and for demonstrating the universality of these phenomena. 
The particles or droplets in such physico-chemical systems are called self-propelled particles.

In terms of symmetry, self-propulsion is classified into two types.
The first type involves the direction of motion being determined by the asymmetry originally embedded in the system, such as geometric~\cite{Nakata, Alcohol} and surface properties~\cite{Howse, Paxton, Sano}.
Several studies on controlling the self-propelled motion have been performed in this regard.
For example, a particle breaking inversion symmetry may exhibit translation~\cite{Nakata, Howse, Paxton, Alcohol, Sano, Kapral}, and one breaking rotational symmetry may exhibit rotation~\cite{Lowen, Nakata, Magome}.
It is also possible to control self-propulsion by geometry of confinement \cite{Ghosh:2013,Fily:2014}.
The second type of the self-propulsion is the motion of a symmetric particle that arises via spontaneous symmetry breaking due to fluctuation~\cite{Sumino1D, Hayashima, Takabatake, Sumino2D, Domingues, Ohta, Buyl, Nagai, Nagayama, Oliver, Ban, Yabunaka:2012, Yoshinaga}.
In this case, the rest state consistent with symmetric properties of the system becomes unstable and motion with lower symmetry is realized. In this article, we focus on symmetric particles.
Oscillatory motion around a system's central position has been observed in many experimental one-dimensional finite systems with inversion symmetry~\cite{Sumino1D, Hayashima}.
We can understand that these oscillatory motions are produced through bifurcation from the rest state at the system center by breaking inversion and time-translation symmetries, though the driving force is invariant under inversion transformation.

As an extension of the one-dimensional system, we consider an isotropic two-dimensional system.
In this system, rotational and oscillatory motions can be taken as self-propelled particle motion, once the rest state is destabilized.
In fact, Takabatake {\it et al.} reported that an oil droplet driven by laser heating via the Marangoni effect exhibits rotational or oscillatory motion depending on the laser intensity although there is no intrinsic asymmetry in the system. 
These authors made a specific model for this phenomenon, which reproduces the experimental results~\cite{Takabatake}.
However, we do not know the generic criteria for how a self-propelled particle is driven toward rotational or oscillatory motion in two-dimensional axisymmetric systems.
Thus, the purpose of this article is to discuss the existence and stability of solutions corresponding to rotation and oscillation.

\section{MODEL}

First, we introduce a model equation. $\bm{r}(t) \in \mathbb{R}^2$ is a vector that denotes the particle position at time $t$. The equation of motion for the particle can be represented as $\ddot{\bm{r}} = \bm{f}(\bm{r},\dot{\bm{r}})$.
This equation can be obtained from the reduction of the mathematical model equations which describe the actual phenomenon in detail. 
$\bm{f}(\bm{r},\dot{\bm{r}})$ is not necessarily real mechanical force but is, rather, an effective force associated with force-free motion for self-propulsion~\cite{Lowen}. 
This description is valid as far as the motion of a single particle is concerned~\cite{footnote}.
In order to discuss bifurcation from the rest state at the system center to rotational or oscillatory motion, we expand $\bm{f}(\bm{r},\dot{\bm{r}})$ with regard to $\bm{r}$ and $\dot{\bm{r}}$ up to the third order near $\bm{r} = \dot{\bm{r}} = 0$. 
We then impose the condition that $\bm{f}(\bm{r},\dot{\bm{r}})$ is invariant under rotation and inversion transformations.
Here, we assume that a self-propelled particle is confined in a two-dimensional parabolic potential around the origin.
After proper nondimensionalization, we obtain a generic equation
\begin{align}
\label{n.d.eq.of.motion}
\ddot{\bm{r}} &= - \bm{r} + b \dot{\bm{r}} + c |\bm{r}|^2 \bm{r} + k |\dot{\bm{r}}|^2 \dot{\bm{r}} \nonumber \\
&\quad + h |\dot{\bm{r}}|^2 \bm{r} + n |\bm{r}|^2 \dot{\bm{r}} + j (\bm{r} \cdot \dot{\bm{r}}) \bm{r} + p (\bm{r} \cdot \dot{\bm{r}}) \dot{\bm{r}}.
\end{align}
It is noted that the coefficients, $b$, $c$, $k$, $h$, $n$, $j$, and $p$, reflect properties of the original model.

Equation (\ref{n.d.eq.of.motion}) is a four-variable dynamical system, which has a trivial fixed point as an origin. The linearized equations around the fixed point are
\begin{align}
\begin{pmatrix}
\dot{x}_1 \\
\dot{v}_1 \\
\dot{x}_2 \\
\dot{v}_2 \\
\end{pmatrix}
=
\begin{pmatrix}
0 & 1 & 0 & 0 \\
-1 & b & 0 & 0 \\
0 & 0 & 0 & 1 \\
0 & 0 & -1 & b
\end{pmatrix}
\begin{pmatrix}
x_1 \\
v_1 \\
x_2 \\
v_2 \\
\end{pmatrix},
\end{align}
where $\bm{r} = {}^\mathrm{t}(x_1, x_2)$ and $\dot{\bm{r}} = {}^\mathrm{t}(v_1, v_2)$.
The eigenequation of the matrix has multiple roots, $b/2 \pm i \sqrt{4 - b^2}/2$, and this system undergoes Hopf bifurcation at $b=0$.
When $b<0$, the rest state at the origin is stable, which means that a particle near the origin eventually has zero displacement. 
To discuss stable rotational and oscillatory motions, we consider cases where $b>0$ in this article.

\section{WEAKLY NONLINEAR ANALYSIS}

We perform weakly nonlinear analysis under the condition that $b \gtrsim 0$ and that cubic terms are on the same order as $b \dot{\bm{r}}$~\cite{Strogatz}.
This condition dictates that the harmonic potential corresponding to the linear restoring force dominates the particle motion.
In this situation, we assume that $\bm{r}(t)$ can be described as,
\begin{align}
\label{r1r2}
\bm{r}(t) = \left (
\begin{array}{l}
x_1 (t) \\
x_2 (t)
\end{array}
\right ) = \left (
\begin{array}{l}
r_1 (\epsilon t) \cos (t + \phi_1(\epsilon t)) \\
r_2 (\epsilon t) \cos (t + \phi_2(\epsilon t))
\end{array}
\right ),
\end{align}
where $r_1$ and $r_2$ denote the amplitudes and $\phi_1$ and $\phi_2$ denote the phases for the oscillation along each axis, which change slowly compared with the harmonic oscillation.
This slow dynamics is owing to the perturbation terms, {\it i.e.}, $b\dot{\bm{r}}$ and cubic terms. Here, the time scale of amplitudes and that of phases are set to be the same, which is defined as $1/\epsilon$ where $\epsilon \ll 1$, in order to include the interaction terms between the amplitudes and the phases in the equations of the slow time-scale variables, $r_1$, $r_2$, $\phi_1$, and $\phi_2$.
Then we substitute eq.~(\ref{r1r2}) into eq.~(\ref{n.d.eq.of.motion}), and obtain the equations for the slow time-scale variables through separating the time scales.
By defining new $2 \pi$-periodic functions $\phi = \phi_1 - \phi_2$ and $\phi_+ = \phi_1 + \phi_2$ in the place of $\phi_1$ and $\phi_2$, we obtain the evolutional equations for $r_1$, $r_2$, $\phi$, and $\phi_+$,
\begin{align}
&\dot{r}_1 = \mu r_1 + A {r_1}^3 + (B + C \cos 2 \phi - D \sin 2 \phi ) r_1 {r_2}^2, \label{r1dot} \\
&\dot{r}_2 = \mu r_2 + A {r_2}^3 + (B + C \cos 2 \phi + D \sin 2 \phi) r_2 {r_1}^2, \label{r2dot}
\\
&\dot{\phi} = D ({r_2}^2 - {r_1}^2) (1 - \cos 2 \phi) - C ({r_1}^2 + {r_2}^2) \sin 2 \phi, \label{phidot}
\\
&\dot{\phi}_+ = -( E + D \cos 2 \phi) ({r_1}^2 + {r_2}^2) - C ({r_2}^2 - {r_1}^2) \sin 2 \phi,
\end{align}
where $\mu = b/2$, $A = (3 k + n + j)/8$, $B = (k + n)/4$, $C = (k - n + j)/8$, $D = (c - h + p)/8$, and $E = (5c + 3h + p)/8$. 
Here $\dot{r_1}$, $\dot{r_2}$, $\dot{\phi}$, and $\dot{\phi}_+$ are functions of only $r_1$, $r_2$, and $\phi$, but independent of $\phi_+$, which means that the present system is intrinsically a three-variable dynamical system on $r_1$, $r_2$, and $\phi$. In this system, rotational and oscillatory motions are represented as fixed points, that is, $(r_1, r_2, \phi) = ( r_{\mathrm{rot}}, r_{\mathrm{rot}}, \pm \pi/2 )$ for rotation and $(r_1, r_2, \phi) = ( r_{\mathrm{osc}} \cos \Psi, r_{\mathrm{osc}} \sin \Psi, 0)$ for oscillation, where $r_{\mathrm{rot}} > 0$, $r_{\mathrm{osc}} > 0$ and $0 \leq \Psi < \pi$.
Here, $\Psi$ represents the angle in the direction of oscillation in the $x_1$-$x_2$ plane.
We therefore obtain the conditions for the existence of these fixed points and then analyze their linear stabilities.

\begin{figure}
\includegraphics{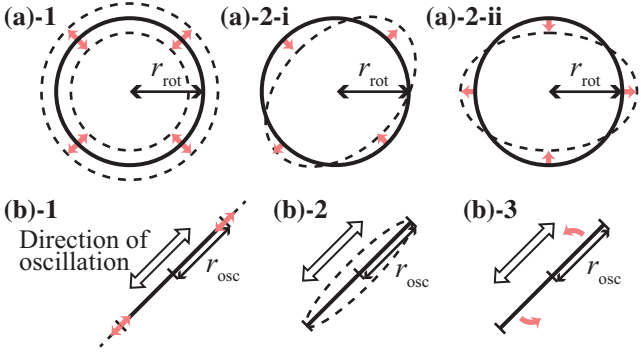}
\caption{Orbit deformations by the eigenvector of each eigenvalue for (a) rotational motion, and (b) oscillatory motion, when $c$, $h$, and $p$ are $0$. 
Red (gray) arrows show the projection of the eigenvectors to the positional plane, and the broken curves show the deformed orbits.
Red (gray) arrows in (a)-1 and (a)-2 correspond to the eigenvectors of eigenvalues $-b$ and $(k-n+j) {r_{\mathrm{rot}}}^2 / 2$, respectively. The orbit deformation for the eigenvalue $-b$ is a radial stretching ((a)-1) and that for the eigenvalue $(k-n+j) {r_{\mathrm{rot}}}^2 / 2$ indicates elliptic deformation ((a)-2). Red (gray) arrows in (b)-1, (b)-2, and (b)-3 correspond to the eigenvectors of eigenvalues $-b$, $-(k-n+j) {r_{\mathrm{osc}}}^2 / 4$, and $0$, respectively. The orbit deformation for the eigenvalue $-b$ is an amplitude stretch ((b)-1), that for the eigenvalue $-(k-n+j) {r_{\mathrm{osc}}}^2 / 4$ is elliptic deformation ((b)-2), and that for the eigenvalue $0$ is rotation of the oscillation direction, $\Psi$ ((b)-3).
}
\label{eigenspace}
\end{figure}

\begin{figure}
\includegraphics{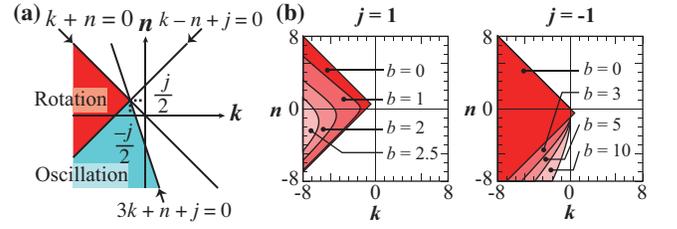}
\caption{(a) Phase diagram on $k$ and $n$ obtained by the weakly nonlinear analysis for fixed $j$. The region with a stable rotational solution is colored with red (dark gray) and that for oscillatory solution with blue (light gray). (b) Phase diagrams on $k$ and $n$ obtained by linear stability analysis of an exact rotational solution for $j=1$ and $-1$. The phase diagrams with $b=0$ in (b) correspond to those in (a) for the same $j$. The colored regions show where a stable rotational solution exists for each $b$.}
\label{pd}
\end{figure}

Firstly, we check for the existence and stability of a solution for rotational motion. By substituting $(r_1, r_2, \phi) = (r_{\mathrm{rot}}, r_{\mathrm{rot}}, \pm \pi/2 )$ into eqs.~(\ref{r1dot})-(\ref{phidot}), and solving $\displaystyle{(\dot{r}_1, \dot{r}_2, \dot{\phi}) = ( 0, 0, 0 )}$, we obtain $r_{\mathrm{rot}} = \displaystyle{ \sqrt{- \mu / 2 B}}$, which occurs only when $4 B = k + n < 0$.
This means that injected energy represented by the term, $b\dot{\bm{r}}$, is dissipated by the summation of the two terms, $k |\dot{\bm{r}}|^2 \dot{\bm{r}}$ and $n |\bm{r}|^2 \dot{\bm{r}}$.
We set $\Delta r_1$, $\Delta r_2$, and $\Delta \phi$ as perturbations around the fixed point. The linearized evolutional equations are then found to be
\begin{align}
\label{wna.linear.rot}
\begin{pmatrix}
\dot{\Delta r_1} \\
\dot{\Delta r_2} \\
\dot{\Delta \phi}
\end{pmatrix}
=
\begin{pmatrix}
\alpha & \beta & \gamma \\ 
\beta & \alpha & -\gamma \\ 
\xi & -\xi & \zeta \\ 
\end{pmatrix}
\begin{pmatrix}
\Delta r_1 \\
\Delta r_2 \\
\Delta \phi
\end{pmatrix}
,
\end{align}
where $\alpha = \mu + 2 (2A - C) {r_{\mathrm{rot}}}^2$, 
$\beta = 2 (A - 2C) {r_{\mathrm{rot}}}^2$, 
$\gamma = 2 D {r_{\mathrm{rot}}}^3$, 
$\xi = - 4 D r_{\mathrm{rot}}$, and 
$\zeta = 4 C {r_{\mathrm{rot}}}^2$.
The eigenvalues of the matrix in eq.~(\ref{wna.linear.rot}) are $-2 \mu (= -b)$ and $4 C {r_\mathrm{rot}}^2 \pm 4 i |D| {r_\mathrm{rot}}^2 (= (k - n + j) {r_\mathrm{rot}}^2/2 \pm i |c-h+p| {r_\mathrm{rot}}^2/2)$, whose eigenvectors are illustrated in Fig.~\ref{eigenspace}(a).
For stability of the fixed point, $8 C = k - n + j< 0$ is required.
Thus, the conditions for a stable solution for rotational motion are summarized as,
\begin{align}
\label{condition.rot}
\left \{
\begin{array}{ll}
4 B = k + n < 0, \\
8 C = k - n + j < 0,
\end{array}
\right .
\end{align}
which are shown in Fig.~\ref{pd}(a).

Secondly, we check for the existence and stability of a solution for oscillatory motion.
By substituting $(r_1, r_2, \phi) = ( r_{\mathrm{osc}} \cos \Psi, r_{\mathrm{osc}} \sin \Psi, 0)$ into eqs.~(\ref{r1dot})-(\ref{phidot}), and solving $\displaystyle{(\dot{r}_1, \dot{r}_2, \dot{\phi}) = ( 0, 0, 0 )}$, we obtain $r_{\mathrm{osc}} = \sqrt{-\mu / A}$, which occurs only when $8 A = 3 k + n + j < 0$.
This means that the summation of the three terms, $k |\dot{\bm{r}}|^2 \dot{\bm{r}}$, $n |\bm{r}|^2 \dot{\bm{r}}$, and $j (\bm{r} \cdot \dot{\bm{r}}) \bm{r}$,  corresponds to energy dissipation just like in the case of rotation.
We set $\Delta r_1$, $\Delta r_2$, and $\Delta \phi$ as perturbations around the fixed point, and the linearized evolutional equations are found to be
\begin{align}
\label{wna.linear.osc}
\begin{pmatrix}
\dot{\Delta r_1} \\
\dot{\Delta r_2} \\
\dot{\Delta \phi}
\end{pmatrix}
=
\begin{pmatrix}
{\alpha}' & {\beta}' & {\gamma}' \\
{\beta}' & {\alpha}'' & {\gamma}'' \\ 
0 & 0 & {\zeta}'
\end{pmatrix}
\begin{pmatrix}
\Delta r_1 \\
\Delta r_2 \\
\Delta \phi
\end{pmatrix},
\end{align}
where ${\alpha}' = - 2 \mu \cos^2 \Psi$, ${\alpha}'' = - 2 \mu \sin^2 \Psi$, 
${\beta}' = - 2 \mu \sin \Psi \cos \Psi$, 
${\gamma}' = - 2 D {r_{\mathrm{osc}}}^3 \sin^2 \Psi \cos \Psi$, 
${\gamma}'' = - 2 D {r_{\mathrm{osc}}}^3 \sin \Psi \cos^2 \Psi$, and 
${\zeta}' = - 2 C {r_{\mathrm{osc}}}^2$.
The eigenvalues of the matrix in eq.~(\ref{wna.linear.osc}) are $-2 \mu (= -b)$, $- 2 C {r_\mathrm{osc}}^2 (= - (k - n + j) {r_\mathrm{osc}}^2 / 4)$, and $0$, whose eigenvectors are illustrated in Fig.~\ref{eigenspace}(b).
For stability of the fixed point, $8C = k - n + j> 0$ is required.
Thus, the conditions for a stable solution for oscillatory motion are summarized as
\begin{align}
\label{condition.osc}
\left \{
\begin{array}{ll}
8A = 3k + n + j < 0, \\
8C = k - n + j > 0,
\end{array}
\right .
\end{align}
which are shown in Fig.~\ref{pd}(a).

From the weakly nonlinear analysis for $b \gtrsim 0$, we can predict which motion occurs (rotation or oscillation), by using the criteria of eqs.~(\ref{condition.rot}) and (\ref{condition.osc}).
It is worth noting that the bistable state, where the rotational and oscillatory motions are both stable, does not exist; this holds as long as $b$ is small enough for the weakly nonlinear analysis to be valid.

In the above discussion, the existence and stability of the solutions for rotational and oscillatory motions are determined by the signs of the linear combinations of coefficients of the three cubic terms, $k |\dot{\bm{r}}|^2 \dot{\bm{r}}$, $n |\bm{r}|^2 \dot{\bm{r}}$, and $j (\bm{r} \cdot \dot{\bm{r}}) \bm{r}$.
Here, we will provide a physical interpretation of these terms.
The term, $k |\dot{\bm{r}}|^2 \dot{\bm{r}}$, denotes velocity-dependent friction, while the terms, $n |\bm{r}|^2 \dot{\bm{r}}$ and $j (\bm{r} \cdot \dot{\bm{r}}) \bm{r}$ denote position-dependent friction.
The latter terms can be rewritten as $n |\bm{r}|^2 \dot{\bm{r}} = n |\bm{r}|^2 (\bm{v}_r + \bm{v}_{\theta})$ and $j (\bm{r} \cdot \dot{\bm{r}}) \bm{r} = j |\bm{r}|^2 \bm{v}_r$. 
Here, $\dot{\bm{r}} = \bm{v}_r + \bm{v}_{\theta}$, where $\bm{v}_r$ is parallel to $\bm{r}$ and $\bm{v}_{\theta}$ is perpendicular to $\bm{r}$.
Thus, $n |\bm{r}|^2 \dot{\bm{r}}$ and $j (\bm{r} \cdot \dot{\bm{r}}) \bm{r}$ are regarded as isotropic and anisotropic friction, respectively.

In contrast with $k$, $n$, and $j$, the coefficients of the other cubic terms, $c |\bm{r}|^2 \bm{r}$, $h |\dot{\bm{r}}|^2 \bm{r}$, and $p (\bm{r} \cdot \dot{\bm{r}}) \dot{\bm{r}}$, do not affect the stability of either rotational or oscillatory motion.
In fact, there is a conserved quantity,
\begin{align}
F(\bm{r}, \dot{\bm{r}}) = \frac 1 2 e^{-w |\bm{r}|^2} \left ( \frac {c}{w^2} + |\dot{\bm{r}}|^2 - \frac {1 - c |\bm{r}|^2}{w} \right ),
\end{align}
where $w = h + p$. Here, we show that $F$ is conserved by calculating the time derivative of $F$ as follows:
\begin{align}
\frac{dF}{dt} &= \sum_{i=1,2} \left ( \frac{dF}{dx_i} \dot{x}_i + \frac{dF}{d v_i} \dot{v}_i \right ) \nonumber \\
&= e^{-w|\bm{r}|^2} \sum_{i=1,2} \left [ - \left ( \frac {c}{w} + w |\dot{\bm{r}}|^2 - 1 + c |\bm{r}|^2 \right ) x_i v_i \right . \nonumber \\
&\quad \left . + \frac {c}{w} x_i v_i + \left ( -x_i + c |\bm{r}|^2 x_i + h |\dot{\bm{r}}|^2 x_i + p (\bm{r} \cdot \dot{\bm{r}}) v_i \right) v_i \right ] \nonumber \\
&= 0,
\end{align}
where we substituted eq.~(\ref{n.d.eq.of.motion}) into $\dot{v}_i$ in the right side of the first line.
This conserved quantity bears a resemblance to energy: when there is no term representing energy injection or dissipation, $b \dot{\bm{r}}$, $k |\dot{\bm{r}}|^2 \dot{\bm{r}}$, $n |\bm{r}|^2 \dot{\bm{r}}$, or $j (\bm{r} \cdot \dot{\bm{r}}) \bm{r}$, $F$ is determined by the initial condition and conserved throughout the orbit.
When $w$ is infinitesimally small, $F$ corresponds to mechanical energy; $F$ is expanded with regard to $w$, and then we obtain
\begin{align}
F(\bm{r}, \dot{\bm{r}}) = \frac{c}{2 w^2} - \frac{1}{2w} + \frac{|\dot{\bm{r}}|^2}{2} + \frac{|\bm{r}|^2}{2} - \frac{c |\bm{r}|^4}{4} + \mathcal{O}(w),
\end{align}
where $\mathcal{O}(w)$ represents the first or higher order terms of $w$. 
The terms, $c/(2 w^2)$ and $- 1/(2w)$, are constant with fixed $w$, and thus $F$ can be considered as summation of kinetic energy, $|\dot{\bm{r}}|^2/2$, and potential energy, $|\bm{r}|^2/2 - c |\bm{r}|^4/4$.
It is noted that the weakly nonlinear analysis discussed above can be regarded as the perturbative approach on the conserved system with sufficiently small energy dissipation/injection terms.

\section{ANALYSIS BEYOND WEAKLY NONLINEAR REGIME}

For rotational motion, we can construct an exact solution for rotational motion when $ {}^\forall b > 0$ and $c = p = h = 0$, since a rotational solution is represented as a fixed point in polar coordinates,
\begin{align}
\label{crd_trs}
\left \{
\begin{array}{ll}
x_1(t) = r(t) \cos \theta(t), \\
x_2(t) = r(t) \sin \theta(t),
\end{array}
\right .
\quad
\left \{
\begin{array}{ll}
v_1(t) = v(t) \cos \psi(t), \\
v_2(t) = v(t) \sin \psi(t).
\end{array}
\right .
\end{align}
By substituting eqs.~(\ref{crd_trs}) into eq.~(1), and by defining new $2 \pi$-periodic functions $\Theta = \psi - \theta$ and $\Theta_+ = \psi + \theta$ in the place of $\psi$ and $\theta$, we obtain
\begin{align}
\dot{r} = v \cos \Theta, \label{rdot_}
\end{align}
\begin{align}
\dot{v} = - r \cos \Theta + b v + \left ( n + \frac j 2 \right ) r^2 v + \frac j 2 r^2 v \cos 2 \Theta + k v^3, \label{vdot}
\end{align}
\begin{align}
\dot{\Theta} = - \frac {v}{r} \sin \Theta + \frac{r}{v} \sin \Theta - \frac j 2 r^2 \sin 2 \Theta, \label{thetadot}
\end{align}
\begin{align}
\dot{\Theta}_+ = \frac {v}{r} \sin \Theta + \frac{r}{v} \sin \Theta - \frac j 2 r^2 \sin 2 \Theta. \label{theta+dot}
\end{align}
Here $\dot{r}$, $\dot{v}$, $\dot{\Theta}$, and $\dot{\Theta}_+$ are functions of $r$, $v$, and $\Theta$, but are independent of $\Theta_+$. Therefore, the dynamical system, eqs.~(\ref{rdot_})-(\ref{theta+dot}), is intrinsically a three-variable system on $r$, $v$, and $\Theta$. In this system, a rotational motion is represented as a fixed point $(r, v, \Theta) = ( r_{\mathrm{rot}}, v_{\mathrm{rot}}, \pm \pi/2 )$. By substituting this into eqs.~(\ref{rdot_})-(\ref{thetadot}), we obtain $r_{\mathrm{rot}} = v_{\mathrm{rot}} = \sqrt{-b/(k+n)}$, which occurs only when $k+n<0$. 
The linearized equations around the fixed point are
\begin{align}
\label{nonweakly-linearizedeq}
\begin{pmatrix}
\dot{\Delta r} \\
\dot{\Delta v} \\
\dot{\Delta \Theta}
\end{pmatrix}
=
\begin{pmatrix}
0 & 0 & -r_\mathrm{rot} \\ 
2 n {r_\mathrm{rot}}^2 & 2 k {r_\mathrm{rot}}^2 & r_\mathrm{rot} \\ 
2/r_\mathrm{rot} & -2/r_\mathrm{rot} & j {r_\mathrm{rot}}^2
\end{pmatrix}
\begin{pmatrix}
\Delta r \\
\Delta v \\
\Delta \Theta
\end{pmatrix}
.
\end{align}
The eigenequation of the matrix in eq.~(\ref{nonweakly-linearizedeq}) is
\begin{align}
\label{eigenequation}
\lambda^3 + \frac{b(2 k + j)}{k+n} \lambda^2 + 2 \left ( \frac{b^2kj}{(k+n)^2} + 2 \right ) \lambda + 4 b = 0.
\end{align}
From the Routh-Hurwitz criterion~\cite{Routh}, we obtain the following conditions for stability of the rotational solution,
\begin{align}
\label{condition.nwna}
\left \{
\begin{array}{l}
2k+j < 0, \\
(2 k + j) k j b^2 + 2 (k - n + j) (k+n)^2 < 0,
\end{array}
\right .
\end{align}
which are shown in Fig.~\ref{pd}(b) for each $b$. For the limit of $b \rightarrow 0$, eq.~(\ref{condition.nwna}) becomes $k - n + j < 0$, which corresponds to the condition for stable rotational motion as shown in eq.~(\ref{condition.rot}) obtained by weakly nonlinear analysis.

Next, we consider the condition of stable oscillation for ${}^\forall b > 0$.
Through weakly nonlinear analysis, we know that the oscillation loses its stability at $3k+n+j=0$ with diverging amplitude and at $k-n+j=0$ with a magnification of the elliptic deformation.
Thus, with $b$ gradually increasing, we predict that the threshold originating from the line, $3k+n+j=0$, can be taken as a problem of a one-dimensional system, $\ddot{x} = - x + b \dot{x} + k \dot{x}^3 + q x^2 \dot{x}$, where $q=n+j$, which was previously investigated by Keith and Rand~\cite{Rand}.
According to their paper, the threshold for the existence and stability of a limit cycle is a line, $3k + q = 0$, when $b \to 0$. The line is bent at $k = q = 0$ as $b$ increases. The degree of bending becomes larger with increasing in $b$, and the threshold approaches a combination of two half lines, $k = 0$ for $q<0$ and $q = 0$ for $k < 0$, when $b \to \infty$.

\section{NUMERICAL CALCULATION}

\begin{figure}
\includegraphics{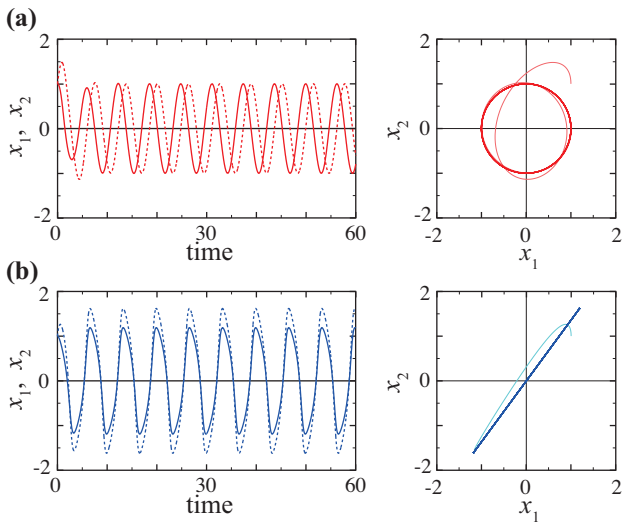}
\caption{Numerical results based on eq.~(\ref{n.d.eq.of.motion}). (a) Results for $k=-1$ and $n=0$. (b) Results for $k=0$ and $n=-1$. The other parameters are $b=1$ and $j=c=h=p=0$. The time series of $x_1$ (solid curves) and $x_2$ (broken curves) are shown in the left panel, and the corresponding trajectories on the $x_1$-$x_2$ plane are shown in the right panel. Both results are calculated with an initial condition, $x_1 = 1$, $x_2 = 1$, $v_1 = 0$, and $v_2 = 1$.
}
\label{orbit}
\end{figure}

We also performed numerical calculation to compare our theoretical results with numerical ones. 
We calculated time evolution based on eq.~(\ref{n.d.eq.of.motion}) using the Euler method with an adaptive time step.
First, we confirmed that eq.~(\ref{n.d.eq.of.motion}) can exhibit stable rotational motion and oscillatory motion as shown in Fig.~\ref{orbit}.
Then we classified the motion into rotation, oscillation and divergence for each $k$ and $n$, and made the phase diagram of the motion in order to compare with the analytical results.

The results are shown in Fig.~\ref{numerical}. The detailed manner to make the phase diagrams is shown in Appendix.
\begin{figure}
\includegraphics{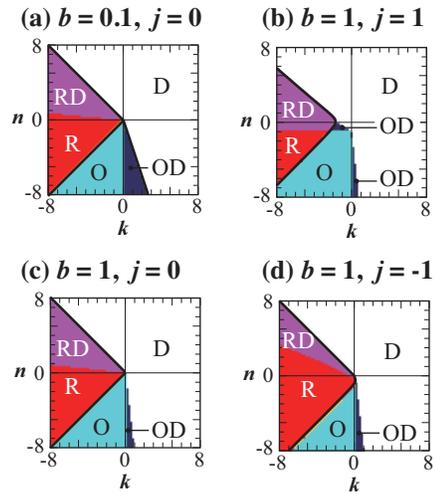}
\caption{Comparison of theoretical and numerical results for (a) $b = 0.1$, and for (b)-(d) $b = 1$. 
The designations, R, O, D, RD, and OD, in the phase diagrams denote rotation, oscillation, divergence, rotation and divergence, and, oscillation and divergence, respectively. 
Results for weakly nonlinear analysis are shown as the black thick lines in (a) and results for the linear stability analysis for the exact rotational solution are shown as the black thick curves in (b)-(d).
}
\label{numerical}
\end{figure}
The numerical results for $b = 0.1$ as shown in Fig.~\ref{numerical}(a) correspond well to the results of the weakly nonlinear analysis, eqs.~(\ref{condition.rot}) and (\ref{condition.osc}).
On the other hand, the numerical results for $b=1$ as shown in Fig.~\ref{numerical}(b)-(d) are in accordance with the results of the linear stability analysis for the exact rotational solution, eq.~(\ref{condition.nwna}).
We also numerically confirmed that the solution structure for the rotational or oscillatory motion barely changes even when the axial symmetry of system is slightly broken.
In the region where $k > 0$ or $n \gtrsim -j$, there are sets of initial values on $x_1$, $x_2$, $v_1$, and $v_2$ from which the trajectory diverges to infinity.
The result in the numerical calculation was not always classified clearly into rotational and oscillatory motion when the parameters are around the boundary between the regions for them.

\begin{figure}
\includegraphics{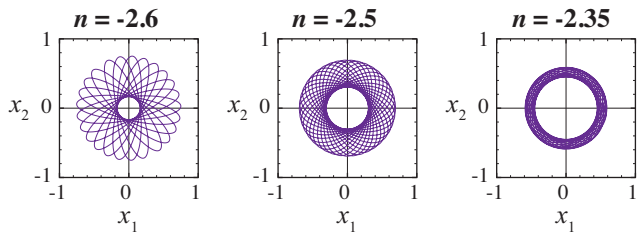}
\caption{Trajectories of quasi-periodic motion on the $x_1$-$x_2$ plane. The short-time trajectories after sufficiently long-time calculation for convergence are shown. The parameters are $b=2$, $j=1$, $k=-5$, and $c=h=p=0$. The value for the parameter $n$ is shown upon each plot.}
\label{qp}
\end{figure}

In order to clarify the detailed structure, we calculated the dependency of motion by precisely scanning the parameter, $n$, across the boundary. 
We set $r_{\mathrm{max}}$ and $r_{\mathrm{min}}$ to be maximum and minimum values of $r$ after sufficiently long-time calculation, respectively.
We characterized the motion by using $\rho = r_\mathrm{min}/r_\mathrm{max}$, where $\rho=1$ and $\rho=0$ correspond to rotational and oscillatory motions, respectively.
\begin{figure}
\includegraphics{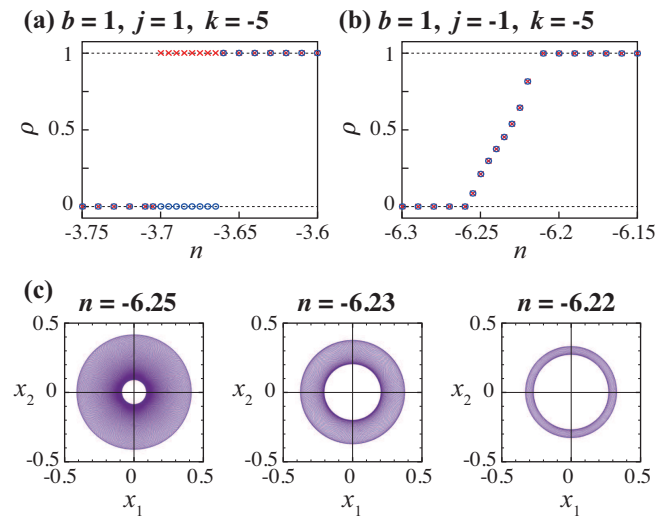}
\caption{Plots of $\rho$ against $n$ with $b=1$, $k=-5$, and (a) $j=1$, (b) $j=-1$. The trajectories were calculated to obtain $\rho$ until time $t \sim 10^6$. 
Crosses and circles represent the results with the initial conditions (i) and (ii) in Supplemental Material~\cite{supporting_1}, respectively.
$\rho=1$, $\rho=0$, and $0<\rho<1$ indicate rotational, oscillatory, and quasi-periodic motion, respectively. The trajectories of the quasi-periodic motion after sufficiently long-time calculation corresponding to (b) are plotted on the $x_1$-$x_2$ plane in (c). The value for the parameter $n$ is shown upon each plot.}
\label{rho}
\end{figure}
Then, we found the parameters where both rotational and oscillatory motions are stable as shown in Fig.~\ref{rho}(a).
There also exist the parameters where the trajectories converge to quasi-periodic orbits. We call such motion as ``quasi-periodic motion'', which is characterized by $\rho$ between 0 and 1. 
Quasi-periodic motion can be seen in a small region sandwiched by the regions for rotational and oscillatory motion as shown in Fig.~\ref{rho}(b).
The trajectories of quasi-periodic motion are shown in Fig.~\ref{rho}(c).
The quasi-periodic orbits are regarded as an elliptic orbit whose long-axis slowly rotates.
They seem to fill the region of $r_{\mathrm{min}} < r < r_{\mathrm{max}}$, if they are drawn over a long period.
For larger $b$, {\it i.e.}, more energy inflow, the angular velocity of rotation of the long-axis becomes larger as shown in Fig.~\ref{qp}, where the shorter-time trajectories are shown.

\section{DISCUSSION AND APPLICATION}

Next, we show several examples where present results can be applied.
First, in a previous study, Sumino {\it et al.} reported a self-propelled droplet exhibiting rotational motion on the surface of a semispherical chamber, and they proposed model equations for the motion of the droplet.
These can be reduced to the following form:
\begin{align}
\label{sumino/reduced}
\ddot{\bm{r}} = \tilde{\mu} \dot{\bm{r}} \left ( {v_0}^2 -\left | \dot{\bm{r}} \right |^2 \right ) + f(|\bm{r}|) \frac{\bm{r}}{|\bm{r}|},
\end{align}
where $\tilde{\mu} > 0$~\cite{Sumino2D}.
When $f(|\bm{r}|) \bm{r} / |\bm{r}|$ can be expanded as a summation of linear restoring force and higher order terms, our criteria, eqs.~(\ref{condition.rot}) and (\ref{condition.osc}), predict that the reduced equation represents rotational motion, since eq.~(\ref{sumino/reduced}) corresponds to the case where $b > 0$, $k < 0$ and $n=j=0$ in eq.~(\ref{n.d.eq.of.motion}).
This agrees with the experimental and numerical results reported in ref.~\cite{Sumino2D}.

Secondly, in a one-dimensional system, the Rayleigh equation~\cite{Rayleigh}, $\ddot{x} + (P_1 + Q_1 \dot{x}^2)\dot{x} + x = 0$, and the van der Pol equation~\cite{vdP}, $\ddot{x} + (P_2 + Q_2 x^2)\dot{x} + x = 0$, are well known as they exhibit limit-cycle oscillation, where $P_1$ and $P_2$ are bifurcation parameters and $Q_1$ and $Q_2$ are positive constants corresponding to energy dissipation.
It should be noted that the Rayleigh equation includes velocity-dependent friction, $\dot{x}^3$, while the van der Pol equation includes position-dependent friction, $x^2\dot{x}$.
We consider the natural extension of these equations to a two-dimensional axisymmetric system:
\begin{align}
\label{Ray.eq}
\ddot{\bm{r}} + (P_1 + Q_1 |\dot{\bm{r}}|^2)\dot{\bm{r}} + \bm{r} = 0, \\
\label{vdP.eq}
\ddot{\bm{r}} + (P_2 + Q_2 |\bm{r}|^2)\dot{\bm{r}} + \bm{r} = 0.
\end{align}
Equation~(\ref{Ray.eq}) exhibits stable rotation, since it corresponds to eq.~(\ref{n.d.eq.of.motion}) with $b > 0$, $k < 0$, and the other coefficients at $0$. 
Meanwhile, eq.~(\ref{vdP.eq}) exhibits stable oscillation, since it corresponds to eq.~(\ref{n.d.eq.of.motion}) with $b > 0$, $n < 0$, and the other coefficients at $0$.
In other words, a self-propelled particle exhibits rotation when friction only depends on its velocity, whereas it exhibits oscillation when friction depends on its position.

In a previous work, Mikhailov and Calenbuhr investigated self-propelled particles in a central force field, and they reported that the particles exhibited rotational motion by using the equation of motion corresponding to eq.~(\ref{Ray.eq})~\cite{Mikhailov}. 
In another previous work, Erdmann {\it et al.} investigated the distribution of active Brownian particles with velocity-dependent friction but without interaction~\cite{Erdmann2000}. They reported that the distribution localizes around a circle at a certain distance from the origin when the particles are in a parabolic potential. When the noise is negligible, the dynamics is almost represented by eq.~(\ref{Ray.eq}) and the particles exhibit rotational motion.

As for the position-dependent friction case, Schweitzer {\it et al.} studied self-propelled particles in a central force field with a localized energy-supplying region; they reported that the particles exhibited oscillatory motion~\cite{Schweitzer, Schweitzer-book}. 
In their work, the location of the energy-supplying region was not symmetric, but their equation roughly corresponds to eq.~(\ref{vdP.eq}).
In a two-dimensional system, therefore, the forms of the dissipation terms are reflected in the mode of motion, {\it i.e.}, rotation or oscillation.

From the viewpoint of the phase dynamics~\cite{Kuramoto}, the proposed model, eq.~(\ref{n.d.eq.of.motion}), can be regarded as coupled limit-cycle oscillators; one is composed of $x_1$ and $v_1$, and the other is composed of $x_2$ and $v_2$, which are coupled with each other by the cubic terms. Then, oscillatory motion in two-dimensional space corresponds to inphase or antiphase synchronization, whereas rotational motion corresponds to the phase-locked state with a phase difference of $\pm \pi/2$. On the contrary, limit-cycle oscillators coupled with each other by linear terms have been widely studied~\cite{Rand1980, Low2003}. In these systems, it is known that inphase or antiphase synchronization is often observed, but the phase-locked state with a phase difference of $\pm \pi/2$ can rarely be realized due to the lack of axisymmetry.
Since the mutual interaction through cubic terms in our model is naturally introduced based on symmetric properties, it may be interesting to analyze our model in terms of coupled oscillators.

\section{SUMMARY}

In summary, we introduced a simple model equation for a self-propelled particle in a two-dimensional axisymmetric system.
In this equation, both rotational and oscillatory motions appear through Hopf bifurcation, but it is nontrivial which motion appears.
By weakly nonlinear analysis, we obtained the criteria, eqs.~(\ref{condition.rot}) and (\ref{condition.osc}), for a self-propelled particle to exhibit rotation or oscillation.
Since we constructed our model equation by assuming only symmetric properties, we believe that our results will contribute widely as a guide for producing reduced mathematical models of self-propelled particles.
It remains as future work to reduce each specific model into the proposed dynamical system and to evaluate the validity of our approach.

\section{Acknowledgements}

We acknowledge Fumi Takabatake for valuable discussion and for sharing data prior to the publication. We also thank Takao Ohta and Yutaka Sumino for helpful discussion. This work was supported in part by Grants-in-aid for Young Scientists (B) to H.K. (No. 24740256) and to N.Y. (No. 26800219), and for Scientific Research on Innovative Areas ``Fluctuation \& Structure'' to H.K. (No. 25103008) and to N.Y. (No. 26103503), the Core-to-Core Program ``Nonequilibrium dynamics of soft matter and information'' to Y.K. and H.K. from the Japan Society for the Promotion of Science (JSPS), and the Cooperative Research Program of ``Network Joint Research Center for Materials and Devices'' to H.K.

\section*{Appendix: Details of Numerical Results in Fig.~\ref{numerical}}

\begin{figure*}[h]
\includegraphics[scale=1.0]{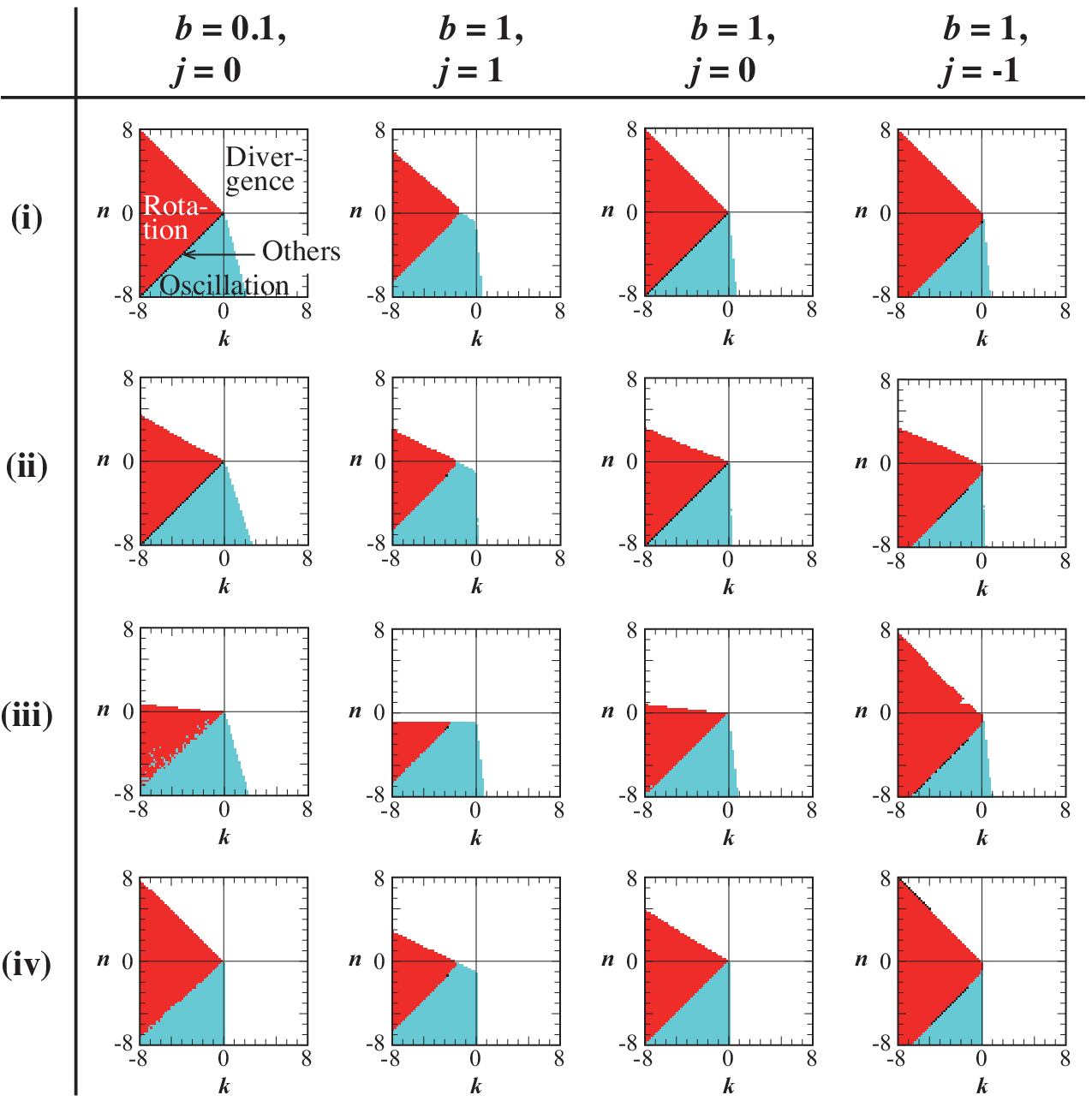}
\caption{Phase diagrams obtained from each initial conditions shown in Table~\ref{ic}. Here, we set $c=h=p=0$. The regions colored with red, blue, white, and dark gray show rotation, oscillation, divergence, and others, respectively. 
}
\label{pd_s}
\end{figure*}
In this appendix, we describe the detailed manner to make the phase diagrams.
We calculated the trajectories for each parameters, and then classified the results as rotation, oscillation, divergence, and others.
To examine the linear stability of the orbit accurately, weak random perturbations are added to $x_1$, $x_2$, $v_1$, and $v_2$ after the trajectory settled to a certain orbit, and then we calculated the trajectory for some time before the classification.
We calculated trajectories from four initial conditions in Table~\ref{ic}.
The initial conditions (i) and (ii) shown in Table~\ref{ic} are set near the orbits estimated from the weakly nonlinear analysis for rotational and oscillatory motions, respectively.
On the while, the initial conditions (iii) and (iv) are set so that the trajectory tends to diverge to infinity based on the analysis of the one-dimensional system (see ref.~\cite{Rand}).
The phase diagrams obtained by numerical calculations from each initial condition are shown in Fig.~\ref{pd_s}. The ambiguous boundaries between rotation and oscillation seen in the phase diagrams for $b=0.1$ with the initial conditions (iii) and (iv), are originated from the randomness of perturbation. Fig.~\ref{numerical} was constructed from the phase diagrams in Fig.~\ref{pd_s} with the same $b$ and $j$.
\begin{table}[h]
\caption{Initial conditions for $x_1$, $x_2$, $v_1$, and $v_2$ used in the numerical calculation. (i)-(iv) correspond to (i)-(iv) in Fig.~\ref{pd}. We define $R_{\mathrm{o}} = \sqrt{\left| 2\mu/(8A+\epsilon) \right|}$, $R_{\mathrm{r}} = \sqrt{\left| 2\mu/(4B+\epsilon) \right|}$, and $K = \sqrt{(n+j)/(k+\epsilon)}$, where $A = (3 k + n + j)/8$, $B = (k + n)/4$, $\mu = b/2$, $\delta = 0.01$, and $\epsilon = 0.005$. 
\label{ic}}
\begin{ruledtabular}
\begin{tabular}{ccccc}
& $x_1$& $x_2$& $v_1$& $v_2$ \\
\hline
(i)& $R_{\mathrm{r}} + \delta$& $\delta$& 0& $R_{\mathrm{r}}$ \\
(ii)& $R_{\mathrm{o}}$& 0& $\delta$& $2\delta$ \\
(iii)& $50 R_{\mathrm{o}}/b$& $50 R_{\mathrm{o}}/b + \delta$& $50 K R_{\mathrm{o}}$& $50 K R_{\mathrm{o}} + \delta$ \\
(iv)& $\delta$& 0& $50 R_{\mathrm{o}}$& $50 R_{\mathrm{o}} + \delta$ \\
\end{tabular}
\end{ruledtabular}
\end{table}

\end{document}